\documentclass[useAMS]{mn2e}
\usepackage{psfig}

\usepackage{times}
\usepackage{amssymb,amsmath}

\def\lsim{ \lower .75ex\hbox{$\sim$} \llap{\raise .27ex \hbox{$<$}} }
\def\gsim{ \lower .75ex \hbox{$\sim$} \llap{\raise .27ex \hbox{$>$}} }

\title[The emission region of PKS B1424--418] 
{The far emission region of the $\gamma$--ray blazar PKS B1424--418}

\author[F. Tavecchio, et al.]
{F. Tavecchio$^1$\thanks{E-mail:fabrizio.tavecchio@brera.inaf.it},
L. Pacciani$^{2,3}$, I. Donnarumma$^{2,3}$, A. Stamerra$^4$, 
J. Isler$^5$, 
E. MacPherson$^5$, 
\newauthor
C. M. Urry$^5$\\
$^1$INAF -- Osservatorio Astronomico di Brera, via E. Bianchi 46, I--23807
Merate, Italy\\
$^2$INAF/IAPS, Via Fosso del Cavaliere 100, I--00133 Roma, Italy\\
$^3$INFN Roma Tor Vergata, I-00133 Roma, Italy\\
$^4$INAF -- Osservatorio Astrofisico di Torino, via P. Giuria 1, I--10125 Torino, Italy\\
$^5$Department of Astronomy, Yale University, PO Box 208101, New Haven, CT 06520-8101, USA.
}

\begin{document}



\maketitle

\begin{abstract} 
We report multi-wavelength (IR-O, UV, X--ray, $\gamma$--ray) data from the flat spectrum radio quasar (FSRQ) PKS B1424-418 ($z=1.52$) gathered during an active  phase in 2013 April . As for a few other cases of FSRQs reported in literature, the detection by the Large Area Telescope onboard {\it Fermi} of $\gamma$ rays with energy above 10 GeV indicates that the emission likely occurs beyond the highly opaque ($\tau\sim10$) broad line region. This conclusion is strengthened by a model fit to the spectral energy distribution, which allows us to locate the emission region even beyond the distance generally assumed for the dusty torus. The consequent large size ($\sim 1$ pc) inferred for the emission region cannot account for the observed daily-scale variability of the $\gamma$--ray flux. We discuss the possibility that  short-term variability results from fast magnetic reconnection events, as proposed in recent works.
\end{abstract}
 

\section{Introduction}

Their favorable geometry makes blazars  the best laboratories for studying relativistic jets. The small angle between the jet velocity vector and the line of sight causes an enormous boost of the relativistically beamed jet non-thermal emission, allowing -- in principle --  the physical processes active in the relativistic outflow to be to probed in detail (e.g., Ghisellini et al. 2008). Currently,  space and ground instrumentation permits an unprecedented view of the spectral energy distribution (SED) and variability of blazars. In particular, the all-sky monitoring by the Large Area Telescope (LAT) onboard {\it Fermi} (Atwood et al. 2009), allows us to follow in detail the behavior of blazars in the $\gamma$--ray (0.1-300 GeV) band, especially the most  powerful flat spectrum radio quasars (FSRQ). Particularly interesting are extreme active phases, during which the large flux  allows us to track precisely the spectral and temporal behavior of the sources.

The classical view of FSRQ assumed that the bulk of the emission occurs in the innermost regions of the active nucleus, within the broad line region (BLR), whose dense radiation field can provide the ideal environment for an efficient inverse Compton (IC) emission, believed to be at the origin of the $\gamma$--ray component (e.g., Sikora et al. 1994). However, there is growing evidence that, at least on some occasions, the emission could occur at much larger distances (up to tens of parsecs; e.g., Marscher et al. 2008, 2010, Sikora et al. 2008). Particularly robust is the indication coming from the detection of $\gamma$ rays with rest-frame energy above 20 GeV (e.g. Aleksic et al. 2011a,b, Abramowski et al. 2013, Pacciani et al. 2012), demonstrating 
that the emission region must be located outside the highly opaque ($\tau \approx$ 5-10) BLR, which would not permit photons of such high energies to escape (e.g., Donea \& Protheroe 2003, Tavecchio \& Mazin 2009, Poutanen \& Stern 2010, but see Tavecchio et al. 2012). This conclusion implies a correspondingly large size of the jet, in contrast with the observed small variability timescale of the $\gamma$--ray flux (from hours down to few minutes), which implies a very compact emission zone. Possibilities to reconcile these two contrasting evidences include the strong recollimation of the jet (Bromberg \& Levinson 2009), narrow beams of electrons (Cerutti et al. 2012) or magnetic reconnection events (Giannios et al. 2009, 2013).

\begin{figure*}
\vspace*{-2.3truecm} 
\hspace{-0.5cm}
\psfig{file=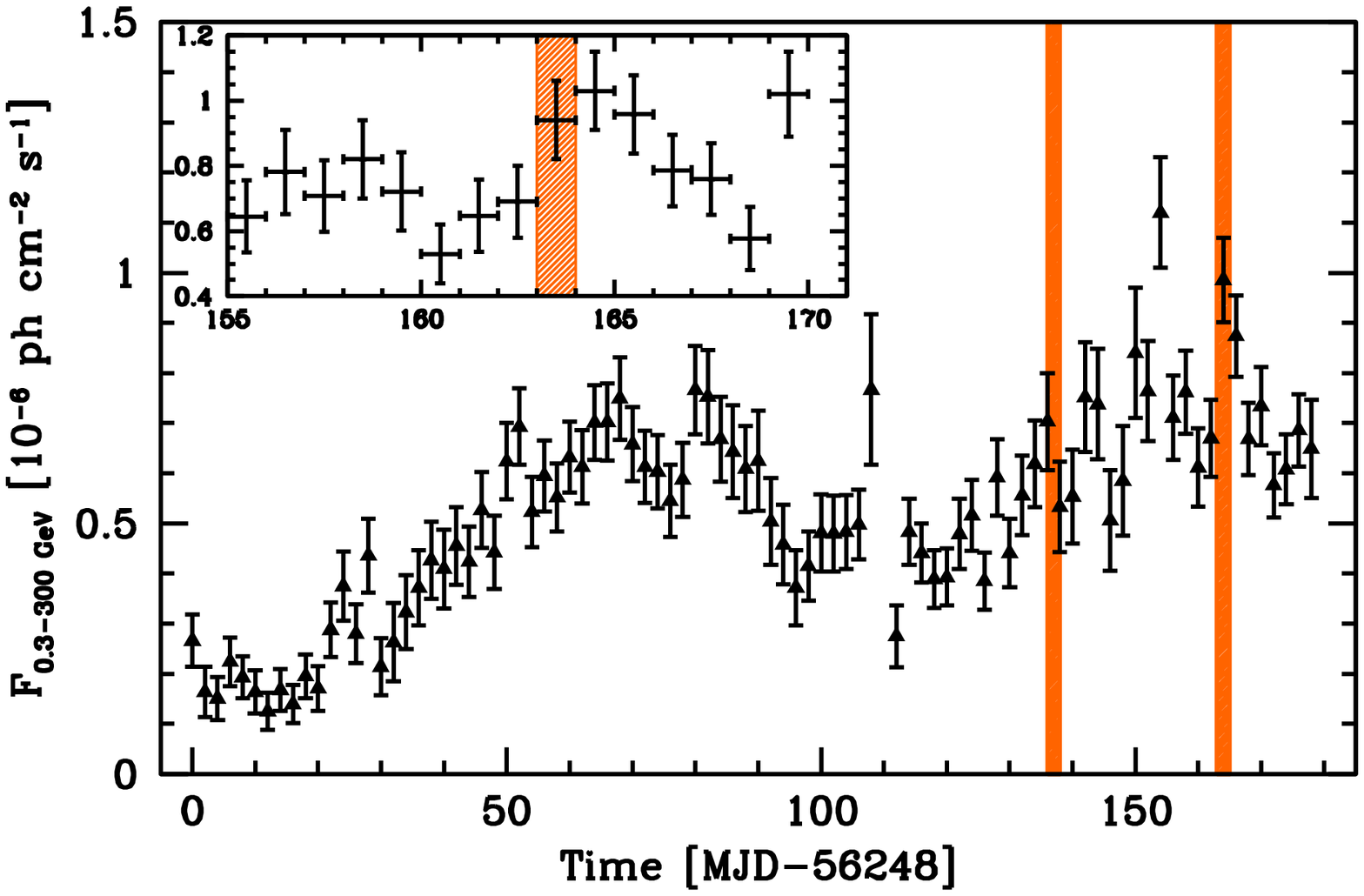,height=12.9cm,width=14.5cm}
\vspace*{-3truecm}
\caption{{\it Fermi}-LAT light curve of PKS B1424-418 in the range 0.3-300 GeV with binning of 2 days. The vertical orange strips correspond to the days (2013 Apr 2 and 29) for which we study the spectral energy distribution. The inset shows a zoom around April 29 (with 1 day bins) to emphasize the daily variability.}
\label{lc}
\end{figure*}

PKS B1424-418 is a relatively bright FSRQ ($z=1.52$) belonging to the classical 1Jy sample and included in the 1FGL and 2FGL catalogues of $\gamma$--ray sources detected by {\it Fermi}-LAT (Nolan et al. 2012). With a notably hard spectrum ($\Gamma=1.96\pm0.03$, to be compared with an average photon index of FSRQ of $\Gamma=2.42\pm 0.17$, Ackermann et al. 2011), this source is a promising candidate for the detection of hard photons. PKS B1424-418 entered in an active $\gamma$--ray phase in  2012 October  (Ojha et al. 2012), with flares at optical, X--ray and $\gamma$--ray frequencies detected in 2013 Jan--Feb  (D'Ammando et al. 2013, Ciprini et al. 2013). It underwent a major $\gamma$--ray outburst in 2013 April, with a peak flux reaching $F(>100 \,{\rm MeV})>3\times 10^{-6}$ ph cm$^{-2}$ s$^{-1}$. A quick analysis of the {\it Fermi}-LAT data revealed the possible presence of hard ($E>10$ GeV) photons. We therefore activated {\it Swift} target of opportunity (ToO) observations to obtain optical-UV and X-ray data. Moreover, PKS B1424-418 was also monitored in the IR-O band with the SMARTS telescopes. In this letter we describe the data obtained in April, focusing on two days (April 2 and 29) with simultaneous  multi wavelength data (\S 2). The composite SED is modeled within the framework of the one-zone leptonic scenario in \S 3 and the results are discussed in \S4.

We assume a cosmology with  $h=\Omega _{\Lambda}=0.7$ and $\Omega _{\rm M}=0.3$. We denote with $c$, $h$ and $k$ the speed of light, the Planck constant and the Boltzmann constant, respectively.

\begin{figure*}
\vspace*{-2.3truecm} 
\hspace{-0.5cm}
\psfig{file=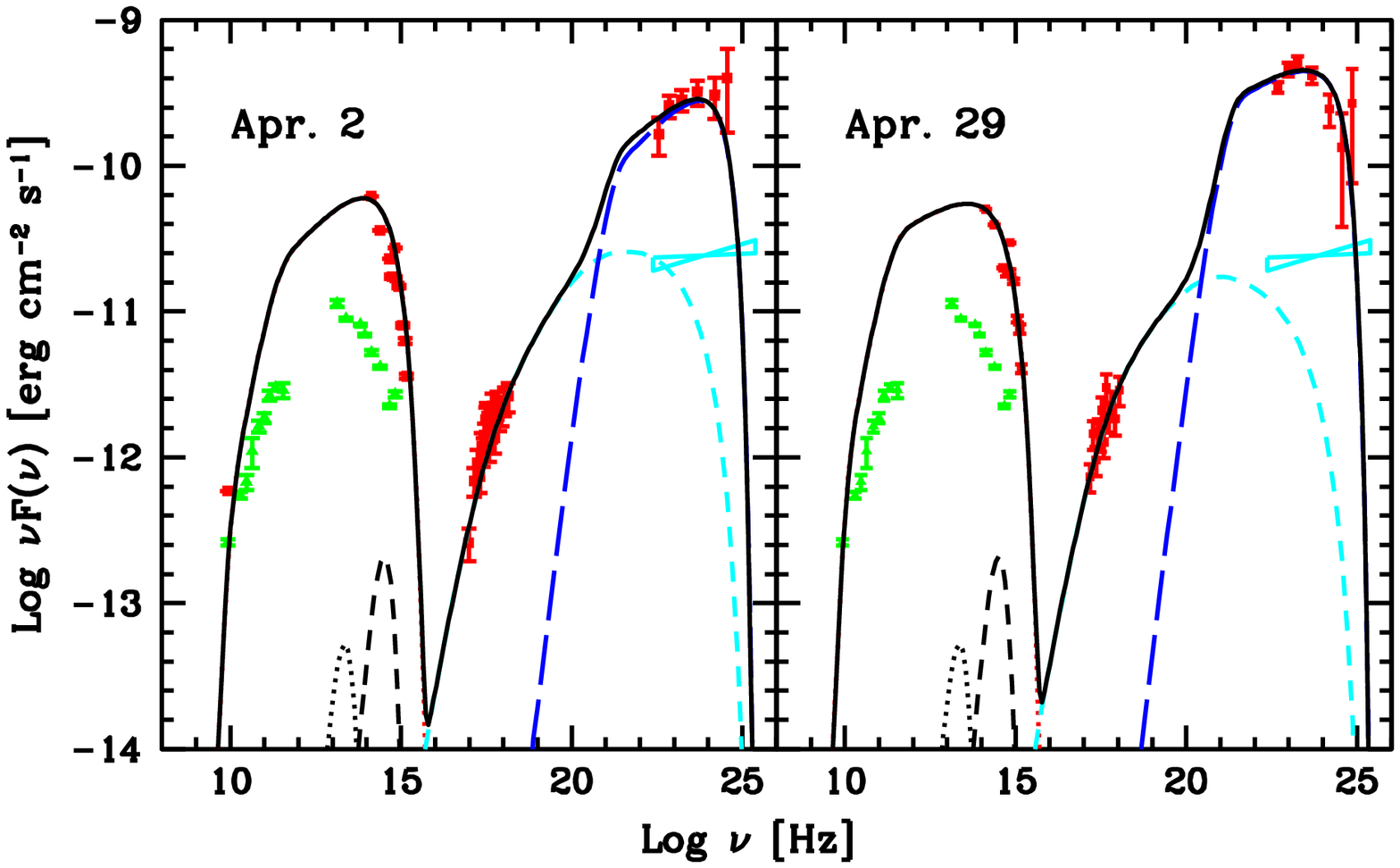,height=13.2cm,width=13.9cm}
\vspace*{-3.15truecm}
\caption{SED of PKS B1424-418 on 2013 Apr. 2 (left) and Apr. 29 (right) built with the data discussed in the text (red points).  Green symbols show archival data from {\it Planck}, ATCA, {\it Wise} and SMARTS data corresponding to a period of low $\gamma$--ray activity (MJD 56069). The cyan butterfly shows the spectrum of PKS B1424-418 in the 2FGL catalogue. The black solid curve reports the emission model (see text for details). We  separately show the SSC (cyan dashed) and EC (long dashed blue) contribution to the total $\gamma$--ray emission emission. The black dashed and dotted lines show the disk and torus emission.}
\label{sed}
\end{figure*}

\section{Data Analysis}

\subsection{{\it Fermi}--LAT data}
We analysed the {\it Fermi}-LAT data (Atwood et al. 2009) with the standard {\it Fermi} Science tools v9r27p1, following the prescriptions in the online documentation\footnote{http://fermi.gsfc.nasa.gov/ssc/data/analysis/documentation/}.
We used the Pass 7 response functions (P7\_V6). In particular, we
selected events of {\em event class} 2, suitable for point--like sources,
and we filtered out photons from the Earth's limb with a cut at
100$^\circ$ in the zenith angle.  We performed the binned likelihood
analysis inside a region of radius 30$^\circ$ around PKS B1424-418 to
derive the source flux.  We took into account the diffuse backgrounds,
(modeled using gal\_2yearp7v6\_v0 for the Galactic diffuse
emission and iso\_p7v6source for the extragalactic isotropic emission
models), Cen A lobes (modeled using the WMAP K--band template emission: CenA\_wmap\_k\_nocenter\_N and CenA\_wmap\_k\_nocenter\_S),
and the  point--like $\gamma$--ray sources in the second {\it Fermi}-LAT catalog (Nolan et al. 2012) within a slightly larger radius of 40$^\circ$ from PKS B1424-418 (we considered a larger radius due to the
PSF width). For each source we used the model specified in the second {\it Fermi}-LAT catalog.
For all the point sources within 10$^\circ$ from PKS  B1424-418, we kept free only the parameters related to the flux normalization, and all other parameters were fixed to the values of the second LAT catalog. 
For the sources outside10$^\circ$ from PKS B1424-418 we fixed all the spectral parameters to the values reported in the catalog.

For the light curves (Fig. \ref{lc})  we consider events only above 300 MeV, in order
to process data with a smaller PSF and then reduce
background $\gamma$ rays from the galactic plane, since
the 68\% (95\%) containment radius for {\it Fermi}-LAT at normal
incidence\footnote{http://www.slac.stanford.edu/exp/glast/groups/canda/archive/lat\_Performance.htm} is 4.5$^\circ$ (10$^\circ$) at 100 MeV.
This choice had a negligible effect on the signal significance of PKS B1424-418. 
The light curves show clear evidence of variability on a timescale of 1-2 days (see zoom in Fig.\ref{lc}). 

SED spectra are integrated over the periods MJD  56381-56385 (left) and MJD  56411-56414 (right).

\subsection{{\it Swift} data}
The \emph{Swift} ToOs that we report here started on 2013-04-02 14:56 and on 2013-04-29 14:41 UT.
The \emph{Swift}--XRT (Burrows et al. 2005) observations, obtained in photon counting mode, 
have an exposure of 3.5~ks and 2.3~ks, respectively.
The XRT data  were processed using
the most recent calibration files available. We made use of
\emph{Swift} software version 3.5, FTOOLS version 6.8, and XSPEC
version 12.5.
We extracted the spectrum using a photon binning ratio that ensured more than 20 photon counts per energy bin.
We fitted the X-ray data with an absorbed power law, fixing the absorption to the Galactic value of 7.7$\times$10$^{20}$ cm$^{-2}$ (Dickey \& Lockman 1990).
For the first observation, we estimated a photon index of 1.60$\pm$0.13  (90\% C.L.) and an unabsorbed flux of 6.2$^{+0.7}_{-0.6}$$\times$10$^{-12}$ erg cm$^{-2}$ s$^{-1}$ (0.3 - 10 keV, 68\% C.L.).
For the second observation, we estimated a photon index of 1.67$\pm$0.20  (90\% C.L.) and an unabsorbed flux of 6.7$^{+0.8}_{-0.7}$$\times$10$^{-12}$ erg cm$^{-2}$ s$^{-1}$ (0.3 - 10 keV, 68\% C.L.).

\textit{Swift}-UVOT (Roming et al. 2005)  data from each observation sequence were
processed by the standard UVOT tool  \texttt{uvotsource} using the
same version of the {\it Swift} software as for the XRT analysis. An
extraction region of radius 5 arcsec centered on the source and a
suitable background region were used. Magnitudes are expressed in the
UVOT photometric system (Poole et al. 2008).
We obtained $m_U$=14.90$\pm$0.03 for the first observation and $m_U$=14.74$\pm$0.03 for the second observation of PKS B1424-418\  (extinction corrected using the mean Galactic interstellar extinction curve from Fitzpatrick 1999).

Both the X-ray and the O-UV light curves provided by the {\it Swift} monitoring between 2013 April 2-29 did not show any significant variation of the flux.

\subsection{SMARTS data reduction}
Observations were carried out with telescopes at Cerro Tololo Interamerican
Observatory (CTIO) operated by the Small and Moderate Aperture Research
Telescope System (SMARTS). The regular monitoring program is carried out
with the SMARTS 1.3m$+$ANDICAM. ANDICAM is a dual--channel imager with a
dichroic that feeds an optical CCD and an IR imager. Thus the instrument
can obtain simultaneous data from 0.4 to 2.2 microns. Observations were
taken nightly in Johnson--Cousins B, R, J, and K bands for PKS B1424-418 from
2013 March 13 through 2013 May 15. Data before that date were obtained on a
three night cadence. Cadences are increased during periods of high
optical and IR flux.

Optical data were bias-- and overscan--subtracted, then flatfielded using
standard IRAF procedures. Infrared data were sky--subtracted, flatfielded,
and dithered images were combined using customized IRAF scripts. Optical
and infrared aperture photometry was also performed using IRAF. We
calibrate comparison stars of comparable magnitude in the target field of
view, using photometric zeropoints that were measured from ANDICAM
observations during 2008--2009 of optical (Landolt 1992) and near--infrared
(Persson 1998) primary standards for each filter. More details
regarding photometric reductions can be found in Bonning et al. (2012).

\section{SED: general constraints and modeling}

{\it General remarks.} Some important and robust insights on the emitting region, independent of a detailed model, can be obtained through rather general arguments. 

\begin{figure}
\vspace*{-0.8cm}
\psfig{file=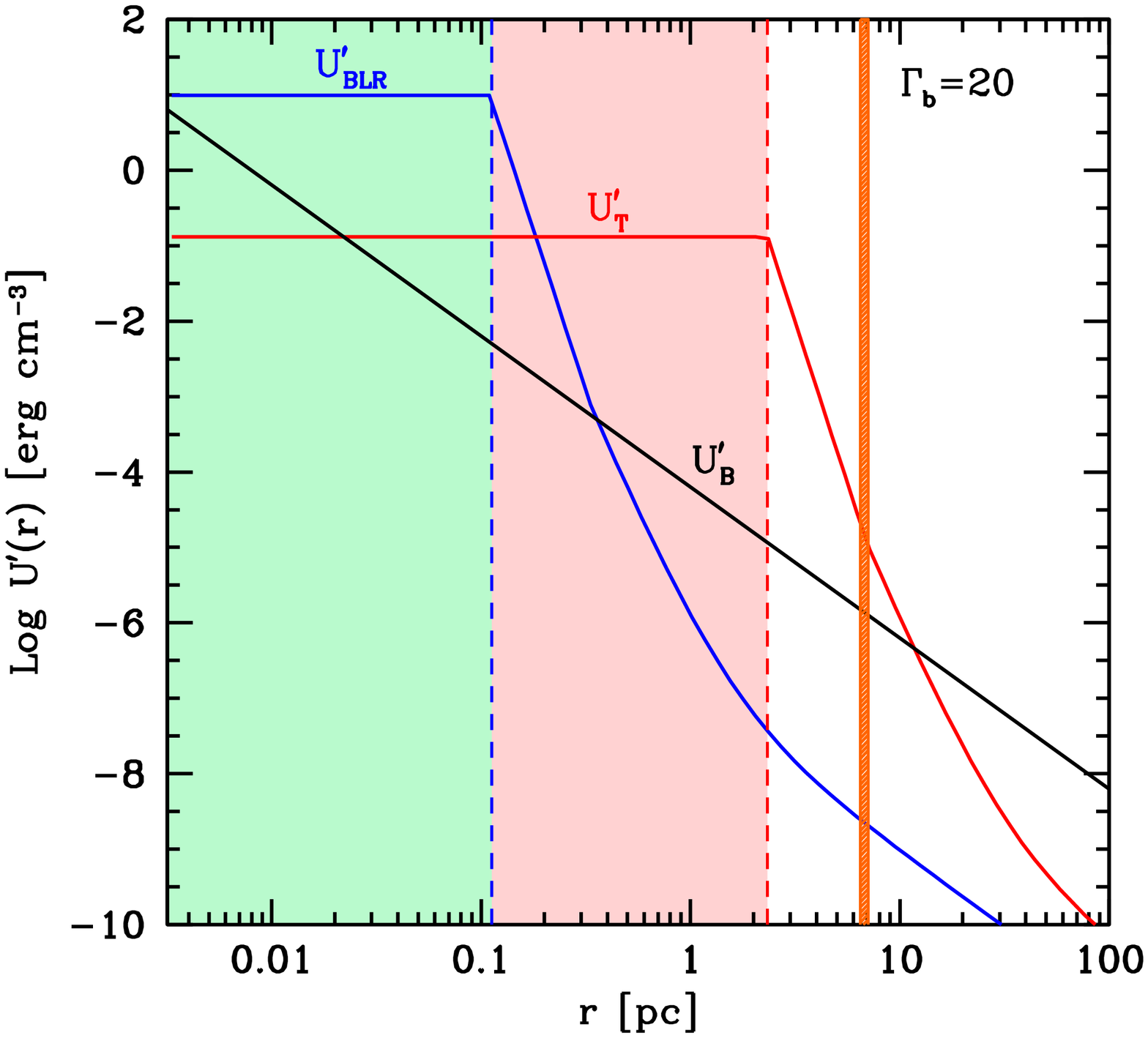,height=7.2cm}
\vspace*{-0.25cm}
\caption{Energy densities in the jet frame (with a bulk Lorentz factor of $\Gamma_{\rm b}=20$) as a function of the distance $r$ along the jet. The BLR and torus energy densities (blue and red lines) are calculated following Ghisellini \& Tavecchio (2009) assuming a disk luminosity $L_{\rm disk}=10^{46}$ erg s$^{-1}$. The black line reports the magnetic energy density $U^{\prime}_B$ normalized to the value derived by the SED fit and assuming a dependence $B^{\prime}\propto r^{-1}$.  The vertical orange line indicates the distance assumed for the models of Fig.\ref{sed}.}
\label{densene}
\end{figure}

The luminosity of the broad MgII line, $L_{\rm MgII}= 5.4\times 10^{43}$ erg s$^{-1}$ (Stickel et al. 1989)\footnote{An  emission line variability study for PKS B1424-418  by the SMARTS group will be addressed in an upcoming paper (Isler et al., in prep).}, can be scaled to the luminosity of the entire BLR using the relative luminosity of the MgII line in quasar spectral templates (we follow the template of Celotti et al. 1997). Assuming a typical BLR/disk luminosity ratio of $\sim$0.1 (e.g. Baldwin \& Netzer 1978), we can also infer the luminosity of the accretion disk, $L_{\rm disk}\simeq10^{46}$ erg s$^{-1}$. With this number at hand, the entire structure of the quasar can be fixed using the prescription of Ghisellini \& Tavecchio (2009) for the BLR and dusty torus sizes, $R_{\rm BLR}$ and $R_{\rm T}$. The result is reported in Fig.\ref{densene}, where the blue and the red lines show the profile of the energy density of the BLR and dusty torus, respectively, as measured in the comoving frame of the jet (assuming a jet bulk Lorentz factor of $\Gamma_{\rm b}=20$). 

A key observational evidence is the possible detection of high-energy photons with energy well above the expected cut-off due to the absorption through interaction with the H Ly$_{\alpha}$ photons within the BLR (e.g., Donea \& Protheroe 2003). 
A simple estimate of the optical depth  close to the energy threshold for interaction with the  H Ly$_{\alpha}$ line ($E_{\rm th}=25$ GeV) provides $\tau\simeq \sigma_{\rm T} \, n_{{\rm Ly}_{\alpha}} \, R_{\rm BLR}/5\simeq 11$, where $\sigma_T$ is the Thomson cross-section. The density of target photons is $n_{{\rm Ly}_{\alpha}}=L_{{\rm Ly}_{\alpha}}/(4\pi R^2_{\rm BLR} c \, h \, \nu _{{\rm Ly}_{\alpha}})$, where $\nu _{{\rm Ly}_{\alpha}}$ is the frequency of the line and, again, the  luminosity in the Ly$_{\alpha}$ line, $L_{{\rm Ly}_{\alpha}}$, is derived following Celotti et al. (1997) from the observed Mg II line.
This result is confirmed by a refined calculation (Tavecchio \& Mazin 2009), showing that the optical depth at the energies of the two highest energy bins observed (15 and 30 GeV; 37.8 and 75.6 GeV in the quasar frame), exceeds $\tau=7$,  implying a suppression factor $F_{\rm obs}/F_{\rm em}=\exp(-\tau)<10^{-3}$. Such a  value excludes an origin for the emission within the BLR. A related, though more model-dependent, argument excluding a dominant UV ambient field is that, in this case, the IC emission (involving the scattering of UV photons) would be suppressed above few GeV due to the onset of the Klein-Nishina regime (e.g., Tavecchio \& Ghisellini 2008). This applies also to the (almost transparent) transition region marked by the BLR edge and the distance at which the IR field starts to dominate. The emission is thus constrained to occur beyond the BLR ``sphere of influence", at distances for which the ambient field is dominated by the IR radiation (Fig. 3).

The SED allows us to derive further clues.
The IC peak occurs at $\nu_{\rm IC}\sim 3\times 10^{24}$ Hz (the most constraining spectrum is that of April 2). The  Lorentz factor of the electrons emitting at the peak, $\gamma ^{\prime}_{\rm p}$, can be evaluated as $\nu_{\rm IC}\simeq \Gamma_{\rm b}^2 \gamma ^{\prime 2}_{p} \nu _{\rm ext}/(1+z)$, where $\nu _{\rm ext}=3.73 \, kT_{\rm IR}/h$ is the frequency of the target photons (with $T_{IR}=10^3$ K) and primed quantities are measured in the jet frame.
Inserting the numerical values we derive $\gamma ^{\prime}_{\rm p}\simeq 10^4 \Gamma _{\rm b, 1}^{-1}$ (where the bulk Lorentz factor is normalized to a typical value of 10, $\Gamma _{\rm b, 1}\equiv \Gamma_{\rm b}/10$ and  we assumed $\Gamma_{\rm b}$ equal to the Doppler factor $\delta$).
From the synchrotron peak frequency, well constrained from above by SMARTS data, $\nu_{\rm s}=2.8\times 10^6 B^{\prime} \, \gamma ^{\prime 2}_{\rm p} \, \Gamma_{\rm b} /(1+z)<3\times 10^{13}$ Hz, we derive an {\it upper limit} to the magnetic field, $B^{\prime}<3\times 10^{-2} \, \Gamma_{\rm b,1}$. This is a value much lower than that commonly inferred for FSRQ,  around 1-10 G (e.g. Ghisellini et al. 2010), consistent with the idea that the region is far from the central engine.  The corresponding  energy density is $U^{\prime}_B<3.2\times 10^{-5}$ erg cm$^{-3} \Gamma_{\rm b,1}^2$. We remark that the unusual value of $B^{\prime}$ is  related to the high IC peak frequency, which forces a large $\gamma ^{\prime}_{\rm p}$, coupled with a low (though unconstrained) synchrotron peak frequency. In the following  we assume the upper value $B^{\prime}=3\times 10^{-2}$ G.

The ratio between the observed IC and synchrotron peak luminosities, $L_{\gamma}/L_{\rm s}\sim 3$, directly translates into the ratio of the external and magnetic energy densities (in the jet frame), allowing the former to be derived: $U_{\rm ext}^{\prime}\approx 10^{-4}$ erg cm$^{-3}$. Such a small value forces us to conclude that the emission region is at a distance of the order of $r\sim 5-10$ pc from the central engine (see Fig.\ref{densene}). We remark that this conclusion is independent of the details of the emission region and it is based solely on rather robust assumptions about the emission mechanisms and the quasar's environment.

A last estimate concerns the radius of the emitting region, for which a lower limit is derived by the condition that the SSC luminosity does not exceed the observed $\gamma$--ray one. This can  be expressed by the condition $L_{\rm SSC}/L_{\rm s}=U^{\prime}_{\rm s}/U^{\prime}_{B}<3$, from which, since $U^{\prime}_{\rm s}=L_{\rm s}/(4\pi R^2 c \, \delta^4)$, we get $R>5\times 10^{17} (\delta/30)^{-4} \Gamma_{\rm b,1}^{-2}$ cm. Assuming a conical geometry of the jet, with semi-aperture  $\theta_{\rm j}\sim 0.1$, this implies a distance $r=R/\theta _{\rm j}>5\times 10^{18}$ cm, consistent with the conclusions above.

\noindent
{\it Modeling the SED.} We model the SED using the one-zone leptonic model  of Maraschi \& Tavecchio (2003). The emission region is assumed to be spherical with radius $R$, in motion with Lorentz factor $\Gamma_{\rm b}$ at an angle $\theta_{\rm v}$ with respect of the line of sight. 
The region contains a tangled magnetic field of intensity $B$ and non-thermal relativistic electrons described by a power-law energy distribution $N(\gamma)=K\gamma^{-n}$ between $\gamma _{\rm min}<\gamma<\gamma _{\rm max}$. The external radiation field in the jet frame (BLR, torus) is modeled using the prescriptions of Ghisellini \& Tavecchio (2009). 

\begin{table*}
 \centering         
       \begin{tabular}{lcccccccccccc}
         \hline
            & $B$ & $K$ & $\gamma _{\rm min}$  & $\gamma_{\rm max}$ & $n$ & $R$ & $\Gamma$ & $r$ & $\theta _{\rm v}$ & $P_{\rm jet}$ & $P_{\rm B}$\\
            & [G] & [cm$^{-3}$] & & & & [cm] & & [cm]& [deg] & [$10^{47}$ erg s$^{-1}]$ & [$10^{44}$ erg s$^{-1}]$\\
         \hline
              Apr 2     & $6\times 10^{-3}$ & $2\times 10^{3}$& $700$&   $4\times10^4$&2.5  & $3.4\times10^{18}$& 20&  $2\times10^{19}$ &1.4 & 1.06& 6.2&\\
              Apr 29   & $6\times 10^{-3}$ & $6.3\times 10^{3}$& $700$&   $4\times10^4$&2.7  & $4.3\times10^{18}$& 20&  $2.15\times10^{19}$ & 1.4 &1.18 & 10.0&\\
          \hline
         \end{tabular}
         \caption{Model parameters used to reproduce the SED in Fig. \ref{sed}. From the first column we report: the magnetic field intensity, the normalization, the minimum and maximum Lorentz factor, the slope of the electron distribution, the radius of the emission region, the bulk Lorentz factor, the distance from the center, the viewing angle, the total jet power (assuming one cold proton per emitting electron) and that carried by  magnetic fields.}
         \label{param}      
         \end{table*}

Fig. \ref{sed} reports the models for the SED of Apr 2 and 29. 
To avoid the overproduction of the X-ray flux, the low energy limit of the electron energy distribution is constrained to be relatively large,  $\gamma _{\rm min}=700$. The EC component only contributes to the high-energy emission, while the SSC emission accounts for the X-rays.

Derived physical parameters are reported in Table \ref{param}. The two different SED have been reproduced changing only the electron distribution ($n$ and $K$) and, slightly, the size and location of the emitting region. In both cases we set the bulk Lorentz factor to the fiducial number $\Gamma_{\rm b}=20$.  The derived SED require a Doppler factor $\delta=32$, from which we derive a viewing angle $\theta_{\rm v}=1.4$ deg. The magnetic field, assumed to scale with the law $B^{\prime}\propto r^{-1}$, is consistent with that expected extrapolating the value (around 1 G) derived for FSRQ in which the emission region is assumed to reside at subpc scales (e.g. Ghisellini et al. 2010).

In Table \ref{param} we report the kinetic (including cold protons, one per electron) and the magnetic power of the jet. 
The value of the power (dominated by the kinetic component) is similar to that typically derived for this class of sources (e.g. Ghisellini et al. 2010). 

\section{Discussion}

The detection of high-energy $\gamma$ rays is a robust {\it model-independent} clue indicating that the emission of PKS B1424-418 during the considered active phase occurred outside the BLR. 

The SED, if interpreted within the leptonic scenario, further allows us to locate the region at few parsecs from the central engine. This conclusion, based solely on the properties of the emission, agrees with other studies in which a similar result has been derived for other blazars through independent evidence (e.g., Marscher et al. 2008, 2010).
The present case is different from that of PMN J2345-1555 (Ghisellini et al. 2013), for which a flare with a hard ($\Gamma<2$) {\it Fermi}-LAT spectrum has been recently reported (however, with no compelling evidence for high-energy $\gamma$ rays.). The qualitative difference between the two cases is that for PMN J2345-1555 the entire SED became ``bluer", with the synchrotron peak also moving to high frequency. This implies a magnetic (and thus an external) energy density not as small as derived for PKS B1424-418, consistent with the emission region being just outside the BLR.

A possibility to avoid the above conclusion would be the absence of the torus itself. In this case the required low value of the external  field (from the BLR alone) could be reached at smaller distances (although the bluer photons would still cause the suppression of the emission above 20 GeV). There is no  evidence of emission from dust in PKS B1424-418; however we expect it to be swamped by the jet non-thermal continuum. Since there are indications for the existence of a ``standard" dusty torus in other $\gamma$--ray blazars (e.g., Malmrose et al. 2011) we consider this possibility unlikely.

The long-term modulation of the light curve (which suggests the presence of two different episodes) is consistent with the inferred light-crossing time, $t_{\rm var}=R/c\delta\sim$ 35 days, and with the radiative cooling timescale, of the order of 30 days for the $\gamma$--ray emitting electrons. However, both timescales are inconsistent with the daily variability. This  problem arises also for other FSRQ for which a far emission region is inferred (the most severe being that of PKS 1222+216, Aleksic et al. 2011). A  way to explain the observations (Tavecchio et al. 2011, Nalewajko et al. 2012) is to admit the existence of very compact regions embedded in the jet, possibly resulting from magnetic reconnection (Giannios et al. 2009, Cerutti et al. 2012). Alternatives include the hadronic  scenario of  Dermer et al. (2012) or the re-collimation of the jet (Bromberg \& Levinson 2009).  We mention in particular the attractive magnetic reconnection scenario of Giannios (2013), which includes in the same scheme a slowly variable  powerful emission (the {\it envelope}) and the rapid flares.  The discussion in Giannios (2013) was suited for PKS 1222+216, for which a smaller distance of the emission region from the central engine is inferred, but modest changes of the parameters should provide a good description also for PKS B1424-418.

Finally, we remark that the conclusions reached for the flare of PKS B1424-418 should not generalized to all blazars. In other cases there are indeed indications favoring an emission region inside the BLR (e.g., Cerruti et al. 2013, Poutanen \& Stern 2010; but see Harris et al. 2012). In any case the investigation of sources for which high-energy ($E>$ 10-20 GeV) $\gamma$ rays can be detected is of paramount importance for a better understanding of the jet physics.

\section*{Acknowledgments}
We are grateful to G. Ghisellini and D. Giannios for discussions. FT acknowledges financial contribution from grant PRIN-INAF-2011. SMARTS observations of LAT monitored blazars are supported by Yale University and Fermi GI grant NNX 12AP15G, and the SMARTS blazar monitoring program is carried out by C.M.U., E.M., Michelle Buxton, Imran Hasan, J. I., Charles Bailyn and Paolo Coppi.
J.I. is supported by the Ford Foundation Dissertation Fellowship. C.B., M.B. and the SMARTS 1.3m observing queue receive support from NSF grant AST-0707627. We thank the {\it Swift} team for the ToO observations. We acknowledge all agencies and Institutes supporting the {\it Fermi}-LAT operations and the  Scientific-analysis tools


\begin{thebibliography}{}

\bibitem[]{} Abramowski, A. et al. 2013, A\&A, in press (arXiv:1304.8071)

\bibitem[\protect\citeauthoryear{Ackermann et al.}{2011}]{2011ApJ...743..171A} Ackermann M., et al., 2011, ApJ, 743, 171 

\bibitem[\protect\citeauthoryear{Aleksi{\'c} et al.}{2011}]{2011A&A...530A...4A} Aleksi{\'c} J., et al., 2011, A\&A, 530, A4 

\bibitem[\protect\citeauthoryear{Aleksi{\'c} et al.}{2011}]{2011ApJ...730L...8A} Aleksi{\'c} J., et al., 2011, ApJ, 730, L8 

\bibitem[\protect\citeauthoryear{Atwood et al.}{2009}]{2009ApJ...697.1071A} Atwood W.~B., et al., 2009, ApJ, 697, 1071 

\bibitem[\protect\citeauthoryear{Baldwin \& Netzer}{1978}]{1978ApJ...226....1B} Baldwin J.~A., Netzer H., 1978, ApJ, 226, 1 


\bibitem[\protect\citeauthoryear{Bonning et al.}{2012}]{2012ApJ...756...13B} Bonning E., et al., 2012, ApJ, 756, 13 

\bibitem[\protect\citeauthoryear{Bromberg \& Levinson}{2009}]{2009ApJ...699.1274B} Bromberg O., Levinson A., 2009, ApJ, 699, 1274 

\bibitem[Burrows et al. (2005)]{burrows2005}Burrows D. N., et al. 2005, Space Sci. Rev., 120, 165

\bibitem[\protect\citeauthoryear{Celotti, Padovani, \& Ghisellini}{1997}]{1997MNRAS.286..415C} Celotti A., Padovani P., Ghisellini G., 1997, MNRAS, 286, 415 

\bibitem[\protect\citeauthoryear{Cerruti et al.}{2013}]{2013arXiv1305.4159C} Cerruti M., Dermer C.~D., Lott B., Boisson  C., Zech A., 2013, ApJ, 771, L4

\bibitem[\protect\citeauthoryear{Cerutti et al.}{2012}]{2012ApJ...754L..33C} Cerutti B., Werner G.~R., Uzdensky D.~A., Begelman M.~C., 2012, ApJ, 754, L33

\bibitem[\protect\citeauthoryear{Ciprini \& Cutini}{2013}]{2013ATel.4770....1C} Ciprini S., Cutini S., 2013, ATel, 4770, 1

\bibitem[\protect\citeauthoryear{D'Ammando et al.}{2013}]{2013ATel.4714....1D} D'Ammando F., Orienti M., Longo F., Jankowsky F., Schwemmer S., Wagn S., 2013, ATel, 4714, 1

\bibitem[\protect\citeauthoryear{Dermer, Murase, \& Takami}{2012}]{2012ApJ...755..147D} Dermer C.~D., Murase K., Takami H., 2012, ApJ, 755, 147 

\bibitem[Dickey \& Lockman (1990)]{dickey1990}Dickey J. M. \& Lockman F. J. , 1990, ARA\&A 28, 215-261

\bibitem[\protect\citeauthoryear{Donea \& Protheroe}{2003}]{2003APh....18..377D} Donea A.-C., Protheroe R.~J., 2003, APh, 18, 377

\bibitem[Fitzpatrick (1999)]{Fitzpatrick1999}Fitzpatrick  E. L., 1999, PASP 111, 63-75


\bibitem[\protect\citeauthoryear{Ghisellini}{2008}]{2008IJMPD..17.1491G} Ghisellini G., 2008, IJMPD, 17, 1491

\bibitem[\protect\citeauthoryear{Ghisellini \& Tavecchio}{2009}]{2009MNRAS.397..985G} Ghisellini G., Tavecchio F., 2009, MNRAS, 397, 985 

\bibitem[\protect\citeauthoryear{Ghisellini et al.}{2010}]{2010MNRAS.402..497G} Ghisellini G., Tavecchio F., Foschini L., 
Ghirlanda G., Maraschi L., Celotti A., 2010, MNRAS, 402, 497 

\bibitem[\protect\citeauthoryear{Ghisellini et al.}{2013}]{2013MNRAS.432L..66G} Ghisellini G., Tavecchio F., Foschini L., 
Bonnoli G., Tagliaferri G., 2013, MNRAS, 432, L66 


\bibitem[\protect\citeauthoryear{Giannios}{2013}]{2013MNRAS.431..355G} Giannios D., 2013, MNRAS, 431, 355 

\bibitem[\protect\citeauthoryear{Giannios, Uzdensky, \& Begelman}{2009}]{2009MNRAS.395L..29G} Giannios D., Uzdensky D.~A., Begelman M.~C., 2009, MNRAS, 395, L29 

\bibitem[\protect\citeauthoryear{Harris, Daniel, \& Chadwick}{2012}]{2012ApJ...761....2H} Harris J., Daniel M.~K., Chadwick P.~M., 2012, ApJ, 761, 2

\bibitem[\protect\citeauthoryear{Landolt}{1992}]{1992AJ....104..340L} Landolt A.~U., 1992, AJ, 104, 340

\bibitem[\protect\citeauthoryear{Malmrose et al.}{2011}]{2011ApJ...732..116M} Malmrose M.~P., Marscher A.~P., Jorstad 
S.~G., Nikutta R., Elitzur M., 2011, ApJ, 732, 116 

\bibitem[\protect\citeauthoryear{Maraschi \& Tavecchio}{2003}]{2003ApJ...593..667M} Maraschi L., Tavecchio F., 2003, ApJ, 593, 667 

\bibitem[\protect\citeauthoryear{Marscher et al.}{2010}]{2010ApJ...710L.126M} Marscher A.~P., et al., 2010, ApJ, 710, L126 

\bibitem[\protect\citeauthoryear{Marscher et al.}{2008}]{2008Natur.452..966M} Marscher A.~P., et al., 2008, Nature, 452, 966 

\bibitem[\protect\citeauthoryear{Nalewajko et al.}{2012}]{2012MNRAS.425.2519N} Nalewajko K., Begelman M.~C., Cerutti B., 
Uzdensky D.~A., Sikora M., 2012, MNRAS, 425, 2519 

\bibitem[Nolan et al. (2012)]{fermicat2}Nolan P. L. et al., 2012, ApJS, 199, 31

\bibitem[\protect\citeauthoryear{Ojha \& Dutka}{2012}]{2012ATel.4494....1O} Ojha R., Dutka M., 2012, ATel, 4494, 1

\bibitem[\protect\citeauthoryear{Pacciani et al.}{2012}]{2012MNRAS.425.2015P} Pacciani L., et al., 2012, MNRAS, 425, 2015 

\bibitem[\protect\citeauthoryear{Persson et al.}{1998}]{1998AJ....116.2475P} Persson S.~E., Murphy D.~C., Krzeminski 
W., Roth M., Rieke M.~J., 1998, AJ, 116, 2475 

\bibitem[Poole et al. (2008)]{Poole08} Poole T. S., Breeveld A. A., Page M. J., et al., 2008, MNRAS, 383, 627

\bibitem[\protect\citeauthoryear{Poutanen \& Stern}{2010}]{2010ApJ...717L.118P} Poutanen J., Stern B., 2010, ApJ, 717, L118 

\bibitem[Roming et al. (2005)]{roming2005}Roming P. W. A., et al. 2005, Space Sci. Rev., 120, 95

\bibitem[\protect\citeauthoryear{Sikora, Begelman, \& Rees}{1994}]{1994ApJ...421..153S} Sikora M., Begelman M.~C., Rees M.~J., 1994, ApJ, 421, 153 

\bibitem[\protect\citeauthoryear{Sikora, Moderski, \& Madejski}{2008}]{2008ApJ...675...71S} Sikora M., Moderski R., Madejski G.~M., 2008, ApJ, 675, 71 

\bibitem[\protect\citeauthoryear{Stickel, Fried, \& Kuehr}{1989}]{1989A&AS...80..103S} Stickel M., Fried J.~W., Kuehr H., 1989, A\&AS, 80, 103 

\bibitem[\protect\citeauthoryear{Tavecchio \& Ghisellini}{2008}]{2008MNRAS.386..945T} Tavecchio F., Ghisellini G., 2008, MNRAS, 386, 945 

\bibitem[\protect\citeauthoryear{Tavecchio \& Mazin}{2009}]{2009MNRAS.392L..40T} Tavecchio F., Mazin D., 2009, MNRAS, 392, L40 

\bibitem[\protect\citeauthoryear{Tavecchio et al.}{2011}]{2011A&A...534A..86T} Tavecchio F., Becerra-Gonzalez J., Ghisellini G., Stamerra A., Bonnoli G., Foschini L., Maraschi L., 2011, A\&A, 534, A86 

\bibitem[\protect\citeauthoryear{Tavecchio et al.}{2012}]{2012PhRvD..86h5036T} Tavecchio F., Roncadelli M., Galanti G., 
Bonnoli G., 2012, Phys. Rev. D, 86, 085036

\end{thebibliography}
\end{document}